\documentclass[aps,twocolumn,amssymb,amsfonts,amsmath,showpacs,final,a4paper]{revtex4}

\usepackage{float}
\usepackage[dvips,final]{graphicx}
\usepackage{color}
\usepackage{amsmath}
\newcommand{\tiem}[1]{{\text{\tiny{#1}}}}
\begin{document}
\title{The Landau-Lifshitz-Bloch equation for ferrimagnetic materials}

\author{U. Atxitia, P. Nieves and O. Chubykalo-Fesenko}
\affiliation{Instituto de Ciencia de Materiales de Madrid, CSIC,
Cantoblanco, 28049 Madrid, Spain}

\date{\today}
%\tableofcontents
\newpage

%##################################################################################################
\begin{abstract}
We derive the Landau-Lifshitz-Bloch (LLB) equation for a two-component magnetic system valid up to the Curie temperature.
As an example, we consider disordered GdFeCo ferrimagnet where the ultrafast optically induced
magnetization switching under the action of heat alone has been recently reported.
The two-component LLB equation contains the longitudinal relaxation terms responding to the
exchange fields  from the  proper and the neighboring sublattices.
We show that the sign of the longitudinal relaxation rate at high temperatures can change depending on the dynamical
magnetization value and a dynamical polarisation of one material by another can occur.
We discuss the differences between the LLB and the Baryakhtar  equation,
recently used to explain the ultrafast switching in ferrimagnets.
The two-component LLB equation forms basis for the large-scale micromagnetic modeling of nanostructures at high temperatures and
ultrashort timescales.
\end{abstract}
\pagebreak
%#################################################################################################
\maketitle

%#################################################################################################
\section{Introduction}
\label{sec:IntroductionLLBferrimagnet}

The Landau-Lifshitz-Bloch (LLB) dynamical equation  of motion for macroscopic magnetization vector \cite{Garanin}
has recommended itself as a valid micromagnetic approach at elevated temperatures \cite{ChubykaloFesenko},
especially useful for temperatures $T$ close to the Curie temperature $T_{C}$ ($T>3T_{C}/4$)
and ultrafast timescales. In several exciting novel magnetic phenomena this approach has been shown to be a necessary tool. These phenomena include laser-induced ultrafast demagnetization
\cite{Atxitia07, Vahaplar, Atxitia10,Sultan}, thermally driven domain wall motion via the spin-Seebeck effect
\cite{Hinzke}, spin-torque effect at elevated temperatures \cite{Haney,Schieback} or heat-assisted magnetic recording \cite{McDaniel}.

In the area of laser-induced ultrafast demagnetization,
the LLB equation has been shown to describe adequately the dynamics in Ni \cite{Atxitia10} and Gd \cite{Sultan}.
The main feature of the LLB equation allowing its suitability for the ultrafast magnetization dynamics is
the presence of longitudinal relaxation term coming from  the strong exchange interaction between atomic spins.
Because the  exchange fields are large ($10-100$ T), the corresponding characteristic longitudinal relaxation timescale
is of the order of 10-100 femtoseconds and thus manifests itself in the ultrafast processes. The predictions of the LLB equations related to the linear reversal path for
the magnetization dynamics \cite{Vahaplar} as well as to the critical slowing down of the relaxation times at high laser pump
fluency \cite{Atxitia10} have been confirmed experimentally.

In ferrimagnetic GdFeCo alloys not only the longitudinal change of magnetization but also a controllable optical magnetization switching has been observed,
and this has stimulated a great deal of effort to
attempt on many levels to explain this process,
see review in Ref. \cite{Kirilyuk}.  The ferrimagnetic
materials consist of at least two antiferromagnetically coupled magnetic sublattices.
The magnetic moments of each sublattice are different, leading to a net
macroscopic magnetization $M(T)$  defined as the sum of magnetization coming from each sublattice.
The main feature of the ferrimagnetic materials is that at some temperature,
called magnetization compensation temperature $T_M$, the
 macroscopic magnetization is zero $M(T_{M})=0$, although   the  magnetization
of each sublattice is not. The angular momentum  compensation temperature, at which the total angular momentum $T_A$ is zero is also of
interest. Simplified considerations of the ferromagnetic resonance of
two-sublattice magnets \cite{Wangsness} predict  that at this temperature the effective damping
is infinite and this stimulated investigation of the magnetization reversal
when going through angular momentum compensation point \cite{StanciuPRB2006,StanciuPRL2007}.

Recently, K. Vahaplar  \emph{et al.} \cite{Vahaplar},
suggested that the  optically induced ultrafast switching in GdFeCo involves a  linear reversal
mechanism, proposed theoretically in Ref. \cite{KazantsevaEPL2009}.
This is an especially fast mechanism since it is governed by the longitudinal relaxation time, which can
be two orders of magnitude faster than the transverse relaxation
time governing precessional switching. The modeling of Ref.\cite{Vahaplar} was based on macrospin LLB approach, essentially treating a ferrimagnet as a ferromagnet.
The model showed that in order to have the magnetization switching a strong field around $20$ T was necessary.
This field can, in principle,  come in the experiment with circularly polarized light from the inverse Faraday effect.
More recently, T. Ostler \emph{et al.} \cite{Ostler} used a multi-spin atomistic approach based
on the Heisenberg model showing that the switching occurs  without any applied field or even with the
field up to $40$ T applied in the opposite direction. The predictions for the heat-driven reversal were confirmed in several experiments in magnetic
 thin films and dots using linearly polarized pulses.  Moreover, I. Radu \emph{et al.} \cite{RaduNATURE2011}
used the same atomistic model for the magnetization dynamics to simulate GdFeCo
and compared the simulation results to the experimental data
measured by the element-specific x-ray magnetic
circular
dichroism (XMCD).
They unexpectedly found that
the ultrafast magnetization reversal in this material, where spins are coupled
antiferromagnetically, occurs by way of a transient ferromagnetic-like
state.

The latter experiments demonstrate the deficiency in application of the macrospin ferromagnetic LLB model to the description
of the ultrafast dynamics in a ferrimagnetic material GdFeCo.
It is clear that the situation of a ferromagnetic-like state in a
ferrimagnetic material cannot be described in terms of a macrospin LLB equation in which a
ferrimagnet is essentially treated as a ferromagnet.
In a ferromagnetic LLB equation the sublattices cannot have their own dynamics and thus the processes such as
the angular momentum transfer between them are essentially ignored. In this situation the only possible
reversal mode is the linear relaxation requiring a strong applied magnetic field as was the case of Ref.\cite{Vahaplar}.

On a general basis, atomistic models are convenient to model ferrimagnetic materials
but for modeling of larger spatial scales, a macroscopic equation similar to ferromagnetic LLB equation
is desirable. This will open a possibility to a correct micromagnetic modeling of ferri- and antiferromagnetic
 nano and micro structures at ultrafast timescales and and/or high temperatures.
Additionally, this can also allow more correct understanding of longitudinal relaxation in two-component (for example, ferrimagnetic) compounds,
taking into account the inter-sublattice exchange.

\begin{figure}[t!]
\centering
\includegraphics[scale=0.8]{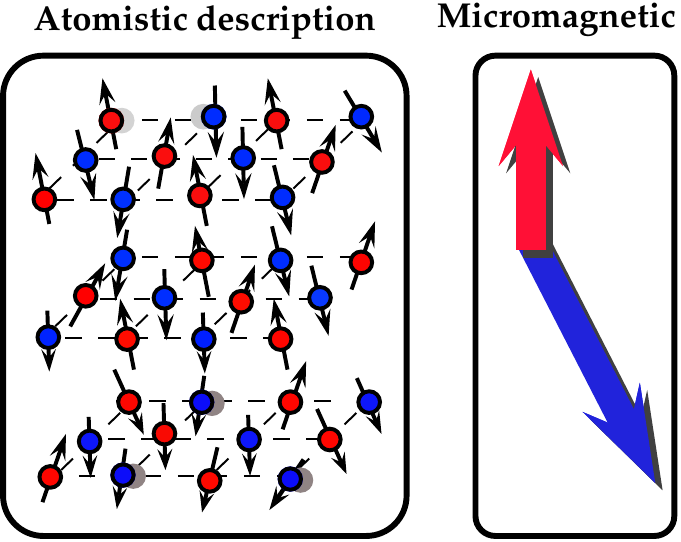}
\caption{ (Left) Sketch of atomistic regular ferrimagnetic lattice. Each point represents a magnetic moment associated with an
atomic site. Magnetic moments of blue points are pointing downwards and red ones upwards.
 (Right) A
macroscopic view of partial average magnetization $m_A=\langle s_A\rangle$ and $m_B=\langle s_B\rangle $
by two macrospins in each sublattice as described by
 the Landau-Lifshitz-Bloch equation.}
\label{fig:ferrimagnet}
\end{figure}

In this article we derive a macroscopic equation for the magnetization dynamics of a
two-component system valid at elevated temperatures in the classical case. As a concrete example, we consider the disordered GdFeCo alloy, the cases of two-component ferromagnets as well as ordered ferrimagnets and antiferromagnets can be easily deduced. Fig.\ref{fig:ferrimagnet} shows a sketch of
an atomistic model for a ferrimagnetic material and the corresponding micromagnetic approximation.
The  atomistic model is based on the classical Heisenberg model for a crystallographically amorphous
ferrimagnetic alloy \cite{OstlerPRB2011} and the Langevin dynamics simulations
 of a set of the Landau-Lifshitz-Gilbert (LLG) equations for localized atomistic spins. In the macroscopic approach each sub-lattice is represented by a macrospin with variable length and direction.
We use the mean field approximation (MFA)  to  derive  a
macroscopic equation of motion for the magnetization of each sublattice.
It contains both transverse and
longitudinal relaxation terms and interpolates between the Landau-Lifshitz equation
at low temperatures and the Bloch equation at high temperatures.
We investigate the signs of the relaxation rates of both transition  (TM) and rare-earth  (RE) metals as a function of temperature.
We conclude that it is a good starting point for performing  large scale simulations in multi-lattice
magnetic systems as the LLB equation is for ferromagnetic materials \cite{Atxitia07,KazantsevaPRB2008}.

\section{ Atomistic model for a disordered ferrimagnet.}
\label{sec:AtomisticModelFerri}

The models for
 binary ferrimagnetic alloys  of the type $A_{x}B_{1-x}$, randomly
occupied by  two different species ($A$ and $B$) of magnetic
ions have been previously extensively investigated theoretically
\cite{MMansuripur1986ieee,MMansuripurBOOK1995,KaneyoshiPRB1986}.
In such models
$A$ and $B$ ions have different atomic quantum spin values $S_{A}$ and
$S_{B}$ ($S_{A}\neq S_{B}$). In the present article we use the
classical counterpart of these models by considering the classical
spins with magnetic moments $\mu_{A}\neq\mu_{B}$. We denote A specie as TM and B specie as RE. A further but non essential simplification
is to assume that the interactions between spins in the disordered
binary alloy are of the Heisenberg form with the exchange interactions
different for different pairs of spins (AA, BB or AB).

Let us start with the
 model for a ferrimagnet described by the classical  Hamiltonian
of the type

%====================== Sketch of a ferrimagnet============
%=========================================================================================
%==================== General impurity Hamiltonian =======================================================
\begin{equation}
\mathcal{H} =  -\sum_{i}^{\mathcal{N}}\mu_{i}\mathbf{H}\cdot\mathbf{s}_{i}
   -\sum_i^{\mathcal{N}} D_i ( s_{i}^{z} )^{2}
  -  \sum_{\langle ij \rangle}J_{ij}\mathbf{s}_{i}\cdot\mathbf{s}_{j},
 \label{eq:binaryalloyGeneralHamiltonian}
 \end{equation}
%=======================================================================================================
where $\mathcal{N}$ is the total  number of spins, $(i,\ j)$
are lattice sites, $\mu_{i}$ is the magnetic moment  located at lattice site $i$.
The external applied field is
 expressed by $\mathbf{H}$.
The anisotropy is considered
as uniaxial with $D_{i}$ being the anisotropy constant of site $i$.
  The third sum is over all nearest and next-to-nearest neighbor pairs and
we have considered unit length classical vectors for all lattice sites $\left|\mathbf{s}_{i}\right|=1$.
Heisenberg exchange interaction parameter between adjacent sites is $J_{ij}=J_{AA (BB)}>0$
if both sites $(i,j)$ are occupied by $A (B)$ type magnetic moments and
$J_{ij}=J_{AB}<0$ if the sites $(i,j)$ are occupied by $A$ and $B$ respectively. We consider that the ordered TM alloy is represented by the fcc-type lattice. To simulate the amorphous character of the TM-RE alloy, $x \cdot 100\%$ lattice sites are substituted randomly with RE magnetic moments.

The magnetization dynamics of this model interacting with the bath is described
by the stochastic Landau-Lifshitz-Gilbert (LLG) equation

%====================================================================================
\begin{equation}
\mathbf{\dot{s}}_{i}=
\gamma_i[\mathbf{s}_{i}\times\mathbf{H}_{i,\textrm{tot}}+\boldsymbol{\zeta}_{i}]-
\gamma_i\lambda_{i}[\mathbf{s}_{i}\times[\mathbf{s}_{i}\times\mathbf{H}_{i,\textrm{tot}}]]
\label{eq:llgdisorder}
\end{equation}
%==========================================================================================
where $\lambda_i$ is the coupling to the heat bath parameter and $\gamma_i$ is the gyromagnetic ratio. In what follows and for simplicity we use the same values for TM and RE, $\gamma_{\textrm{\tiem{TM}}}=\gamma_{\textrm{\tiem{RE}}}=\gamma=1.76 \cdot 10^7 $rad s$^{-1}$ Oe$^{-1}$, $\lambda_{\textrm{\tiem{TM}}}=\lambda_{\textrm{\tiem{RE}}}=\lambda=0.1$. The stochastic thermal fields $\boldsymbol{\zeta}_{i}$ are
uncorrelated in time and on different lattice sites. They can be coupled to
different heat baths (via temperature of phonon or electron) and could have different strength of coupling
(via $\lambda_{i}$ and $\mu_i$) for each atom type ($A$ or $B$).
The correlators of different components of thermal field can be written as:

%===================================================================================================
\begin{equation}
\left\langle \zeta_{i,\alpha}(t)\zeta_{j,\beta}(t')\right\rangle =\frac{2{\lambda}_{i}k_{\text{B}}T}{{\mu}_{{i}}{\gamma}_{i}}\delta_{ij}\delta_{\alpha\beta}\delta(t-t')
\label{eq:white_noise_correlator}
\end{equation}
%==================================================================================================
where $\alpha,\beta$ are Cartesian components and $T$ is the temperature of the
heat bath to which the spins are coupled. The effective fields are given by

%==============================================================================================================0
\begin{equation}
\mathbf{H}_{i,\textrm{tot}}^{\textrm{}}\equiv-\frac{1}{\mu_{i}}\frac{\partial\mathcal{H}}{\partial\mathbf{s}_{i}}=
\mathbf{H}+\frac{2D_i}{\mu_i}s_i^z \mathbf{e}_z+\frac{1}{\mu_{i}}\sum_{j\in\text{neig}(i)}J_{ij}\mathbf{s}_{ij}
\nonumber
\label{eq:Hitotfield}
\end{equation}
%=====================================================================================================================
The particular values for exchange parameters  and the anisotropy
constants (see Table \ref{table_parameters}) are chosen
in such a way that the static properties coincide with experimental measurements
in GdFeCo \cite{OstlerPRB2011}.

%============================================================================================================0
\begin{table}[h!]\label{1}
	\begin{center}
\begin{tabular}{ccccc}
\hline \hline
  & $\mu/\mu_B $ & $D \,\mbox{[Joule]}$ & $J\,\mbox{[Joule]}$  \\ \hline
Transition Metal (TM) & $2.217$ & $8.0725\times10^{-24}$  & $4.5\times 10^{-21}$  \\
Rare-Earth (RE) & $7.63$ & $8.0725\times10^{-24}$  & $1.26\times 10^{-21}$  \\
TM-RE & $-$ & $-$ & $-1.09\times 10^{-21}$  \\
\hline \hline
\end{tabular}
\caption{Table with parameters of transition metal (TM) and rare-earth (RE) compounds.
Anisotropy constant $D_{\tiem{TM(RE)}}$ is taken equal for both lattices.
Exchange parameters $J_{\tiem{TM(RE)}}$/per link  are taken in order to give correct Curie temperature
of pure compounds ($x=0$ pure TM or $x=1$ pure RE).
Antiferromagnetic exchange parameter $J_{\tiem{RE-TM}}$ is  chosen so that the
temperature dependence of the TM and RE sublattices agrees
qualitatively with results of XMCD measurements of static
magnetization \cite{OstlerPRB2011}. }
\label{table_parameters}
\end{center}
\end{table}
%================================================================================

\section{LLB equation for classical ferrimagnet}
\subsection{Equation derivation}
The idea of the two-component LLB model is presented in Fig. \ref{fig:ferrimagnet}.
Namely, our aim is to evaluate the dynamics of the macrosopic classical polarization $\mathbf{m}=\left\langle \mathbf{s}\right\rangle^{\textrm{\tiem{conf}}}$,
where the average is performed over temperature as well as the microscopic disorder configurations.

The dynamics of the mean magnetization can be obtained through the Fokker-Planck equation (FPE)
for non-interacting spins \cite{Garanin}.
%From here the dynamical equation for the spin average can be obtained \cite{RiskenBOOK1989}.
%==============================================================================================================0
%\begin{equation}
%\frac{d\mathbf{m}}{dt}=\gamma[\mathbf{m}\times\mathbf{H}]-\Lambda_{N}\mathbf{m}-\gamma\lambda\left\langle[\mathbf{s}\times[\mathbf{s}\times\mathbf{H}]]\right\rangle
%\label{eq:Fokker-Planck}
%\end{equation}
%=====================================================================================================================
The FPE for the distribution function of an ensemble of interacting spins can be derived
in the same way as in the ferromagnetic case \cite{Garanin}. The FPE
 has as the static solution the Boltzmann distribution function
$f_{0}\left(\left\{ \mathbf{s}_{i}\right\} \right)\propto\exp\left[-\beta\mathcal{H}\left(\left\{ \mathbf{s}_{i}\right\} \right)\right]$,
where $\mathcal{H}$ is given by Eq. \eqref{eq:binaryalloyGeneralHamiltonian} and $\beta=1/(k_B T)$.
Since the exact solution is impossible even in the simple ferromagnetic
case, then,  we  resort to the mean field approximation (MFA)
with respect to spin-spin interactions and random average with respect
to disorder configurations. In the MFA the distribution function is multiplicative
and we can use the same strategy as in the ferromagnetic case  \cite{Garanin}, we
take the distribution function $f_i$ of each lattice site $i$, which satisfy the FPE for a non-interacting spin
 and perform the substitution  $\mathbf{H}\Rightarrow\left\langle \mathbf{H}_{\nu}^{\text{MFA}}\right\rangle^{\textrm{\tiem{conf}}}$,
where $\nu=$TM or RE indicates the sublattices.
Thus, we start with the paramagnetic LLB equation which was derived in the original article by D. Garanin \cite{Garanin} and
is equally valid for the present purpose and substitute the external field by the MFA one in each sublattice.
The  corresponding set of coupled LLB equations for each sublattice magnetization
$\mathbf{m}_{\nu}$ has the following form:

%================================================================================================0
\begin{eqnarray}
\mathbf{\dot{m}}_{\nu} & =&  \gamma_{\nu}[\mathbf{m}_{\nu}\times\left\langle \mathbf{H}_{\nu}^{\text{MFA}}\right\rangle^{\textrm{\tiem{conf}}}]-
\Gamma_{\nu,\|}\left(1-\frac{\mathbf{m}_{\nu}\mathbf{m}_{0,\nu}}{m_{\nu}^{2}}\right)\mathbf{m}_{\nu}
 \nonumber\\
  &-&  \Gamma_{\nu,\bot}
  \frac{[\mathbf{m}_{\nu}\times[\mathbf{m}_{\nu}\times\mathbf{m}_{0,\nu}]]}{m_{\nu}^{2}},
 \label{eq:LLBequationFerrimagnet}
 \end{eqnarray}
%===============================================================================================
where

%===========================================================================================
\begin{equation}
\mathbf{m}_{0,\nu}=B(\xi_{0,\nu})\frac{\boldsymbol{\xi}_{0,\nu}}{\xi_{0,\nu}},
\ \ \boldsymbol{\xi}_{0,\nu}\equiv\beta\mu_{\nu}\left\langle \mathbf{H}_{\nu}^{\text{MFA}}\right\rangle^{\textrm{\tiem{conf}}}.
\label{eq:m0xi0}
\end{equation}
%===========================================================================================
Here $\xi_{0,\nu}\equiv\left|\boldsymbol{\xi}_{0,\nu}\right|$,
 $B\left(\xi\right)=\coth\left(\xi\right)-1/\xi$ is the Langevin
function,
%================================================================================================
\begin{equation}
 \Gamma_{\nu,\|}=\Lambda_{\nu,N}\frac{B(\xi_{0,\nu})}{\xi_{0,\nu}B'(\xi_{0,\nu})},\quad \Gamma_{\nu,\bot}=\frac{\Lambda_{\nu,N}}{2}\left(\frac{\xi_{0,\nu}}{B(\xi_{0,\nu})}-1\right)
\label{eq:relaxations}
\end{equation}
%================================================================================================
describe parallel and perpendicular relaxation, respectively, $\Lambda_{\nu,N}=2\gamma_{\nu}\lambda_{\nu}/\beta\mu_{\nu}$
is the characteristic diffusion relaxation
rate or, for the thermo-activation escape problem, the
N\'{e}el attempt frequency.

Next step is to use in Eqs. \eqref{eq:LLBequationFerrimagnet} and \eqref{eq:m0xi0} the MFA expressions.
The MFA treatment for the disordered ferrimagnet has been presented in Ref. \cite{{OstlerPRB2011}}.
The resulting expressions for the fields have the following forms:

%=================================================================================================
\begin{equation}
\langle\mathbf{H}_{\textrm{\tiem{RE}}}^{\textrm{\tiem{MFA}}}\rangle^{\textrm{\tiem{conf}}}
=\mathbf{H}'_{\textrm{eff,\tiem{RE}}}+\frac{J_{0,\textrm{\tiem{RE}}}}{\mu_{\textrm{\tiem{RE}}}}
\mathbf{m}_{\textrm{\tiem{RE}}}
+\frac{J_{0,\textrm{\tiem{RE-TM}}}}{\mu_{\textrm{\tiem{RE}}}}\mathbf{m}_{\textrm{\tiem{TM}}}
\label{eq:HmfaRE}
\end{equation}
%===================================================================================================
%=======================================================================================================
\begin{equation}
\langle\mathbf{H}_{\textrm{\tiem{TM}}}^{\textrm{\tiem{MFA}}}\rangle^{\textrm{\tiem{conf}}}
=\mathbf{H}'_{\textrm{eff,\tiem{TM}}}+\frac{J_{0,\textrm{\tiem{TM}}}}{\mu_{\textrm{\tiem{TM}}}}\mathbf{m}_{\textrm{\tiem{TM}}}
+\frac{J_{0,\textrm{\tiem{TM-RE}}}}{\mu_{\textrm{\tiem{TM}}}}\mathbf{m}_{\textrm{\tiem{RE}}}
\label{eq:HmfaTM}
\end{equation}
%======================================================================================================
where  $J_{0,\textrm{\tiem{TM}}}=qzJ_{\textrm{\tiem{TM-TM}}}$,
$J_{0,\textrm{\tiem{RE}}}=xzJ_{\textrm{\tiem{TM-TM}}}$,
$J_{0,\textrm{\tiem{RE-TM}}}=qzJ_{\textrm{\tiem{TM-RE}}}$,
$J_{0,\textrm{\tiem{TM-RE}}}=xzJ_{\textrm{\tiem{TM-RE}}}$,
$z$ is the number of nearest neighbors between TM moments in the ordered lattice,  $x$ and $q=1-x$ are the RE and TM concentrations. The field $\mathbf{H}'_{\textrm{eff},\nu}$  contains the external applied and the anisotropy fields acting on the sublattice $\nu =$TM,RE.

The equilibrium magnetization of each sublattice $m_{e,\nu}$ within the MFA approach can be obtained
via the self-consistent solution of the Curie-Weiss equations
%=========================================================================================================
\begin{equation}
\mathbf{m}_{\textrm{\tiem{RE}}}=B\left(\xi_{\textrm{\tiem{RE}}}
\right)\frac{\boldsymbol{\xi}_{\textrm{\tiem{RE}}}}
{\xi_{\textrm{\tiem{RE}}}}
;\quad\mathbf{m}_{\textrm{\tiem{TM}}}=
B\left(\xi_{\textrm{\tiem{TM}}}\right)\frac{\boldsymbol{\xi}_{\textrm{\tiem{TM}}}}{\xi_{\textrm{\tiem{TM}}}}.
\label{eq:CurieWeissImp}
\end{equation}
%===========================================================================================
The resulting equation (\ref{eq:LLBequationFerrimagnet}) with expressions (\ref{eq:HmfaRE})
and (\ref{eq:HmfaTM})  constitutes the LLB equation for a ferrimagnet and
can be already used for numerical modeling at large scale since in what follows
some approximations will be used.
The use of these approximations is necessary for understanding the relaxation
of a ferrimagnetic system from theoretical point of view. We will also get the LLB equation in a more explicit and compact form.

We treat the most general case
where the continuous approximation in each sub-lattice can be used.
Basically, in the spirit of the MFA approximation, in each sub-lattice we treat the $k=0$ mode.
In order to handle the problem analytically we decompose the magnetization vector
$\mathbf{m}{}_{\nu}$
into two components $\mathbf{m}_{\nu}=\boldsymbol{\Pi}_{\nu}+\boldsymbol{\tau}_{\nu}$,
where $\boldsymbol{\Pi}_{\nu}$ is perpendicular to $\mathbf{m}{}_{\kappa}$,
so that it can be expressed as $\boldsymbol{\Pi}_{\nu}=
-\left[ \mathbf{m}_{\kappa}\times\left[\mathbf{m}_{\kappa}
\times\mathbf{m}_{\nu}\right]\right]/m^ 2_{\kappa}$,
and $\boldsymbol{\tau}_{\nu}$ is parallel to $\mathbf{m}_{\kappa}$,
and it can be expressed as $\boldsymbol{\tau}_{\nu}=\mathbf{m}_{\kappa}\left(\mathbf{m}_{\nu}\cdot\mathbf{m}_{\kappa}\right)/m_{\kappa}^{2}$,
where  $\kappa\neq\nu$.

 We can shorten the notation by definition
of the following new variable $\Theta_{\nu\kappa}$

%==========================================================================================
\begin{equation}
\Theta_{\nu\kappa}=\frac{\mathbf{m}_{\nu}\cdot\mathbf{m}_{\kappa}}{m_{\kappa}^{2}}
\Longrightarrow
\mathbf{m}_{\nu}=
\boldsymbol{\Pi}_{\nu}+\Theta_{\nu\kappa}\mathbf{m}_{\kappa}.
\end{equation}
%=========================================================================================00
 As a consequence, the MFA exchange field $\langle\mathbf{H}_{\textrm{EX},\nu}^
{\textrm{\tiem{MFA}}}\rangle^{\textrm{\tiem{conf}}}$  in Eqs.  (\ref{eq:HmfaRE}) and (\ref{eq:HmfaTM})
can be written as the sum of the exchange
fields parallel and perpendicular to magnetization of the sublattice $\nu$.

%===============================================================================0
\begin{eqnarray}
\langle\mathbf{H}_{\textrm{EX},\nu}^
{\textrm{\tiem{MFA}}}\rangle^{\textrm{\tiem{conf}}} & = & \left(\frac{J_{0,\nu}}{\mu_{\nu}}+\frac{J_{0,\nu\kappa}}{\mu_{\nu}}\Theta_{\kappa\nu}\right)\mathbf{m}_{\nu}+\frac{J_{0,\nu\kappa}}{\mu_{\nu}}\boldsymbol{\Pi}_{\kappa}\nonumber\\
 & = & \frac{\widetilde{J}_{0,\nu}}{\mu_{\nu}}\mathbf{m}_{\nu}+\frac{J_{0,\nu\kappa}}{\mu_{\nu}}\boldsymbol{\Pi}_{\kappa}\nonumber\\
 & = & \mathbf{H}_{\textrm{EX},\nu}^{\Vert}+\mathbf{H}_{\textrm{EX},\nu}^{\bot}\label{eq:Heff}
 \end{eqnarray}
%========================================================================================
where we have defined a new function $\widetilde{J}_{0,\nu}\left(\mathbf{m}_{\kappa},\mathbf{m}_{\nu}\right)$
as $\widetilde{J}_{0,\nu}=J_{0,\nu}+J_{0,\nu\kappa}\Theta_{\kappa\nu}\left(\mathbf{m}_{\kappa},\mathbf{m}_{\nu}\right)$,
we remark that $\widetilde{J}_{0,\nu}$ is not a constant but a
function of both sublattice magnetization.
The exchange field is, therefore, separated in two contributions,
a longitudinal one $\mathbf{H}_{\textrm{EX},\nu}^{\Vert}=(\widetilde{J}_{0,\nu}/\mu_{\nu})\mathbf{m}_{\nu}$
and a transverse one $\mathbf{H}_{\textrm{EX},\nu}^{\bot}=(J_{0,\nu\kappa}/\mu_{\nu})\boldsymbol{\Pi}_{\kappa}$.

In the following we will consider that the transverse contribution is small in comparison to longitudinal one, \emph{i.e.}
$|\mathbf{H}_{\textrm{EX},\nu}^{\Vert}|\gg | \mathbf{H}_{\textrm{EX},\nu}^{\bot}  |$.
Finally, $\left\langle \mathbf{H}_{\nu}^{\text{MFA}}\right\rangle^{\text{conf}}  \simeq  \mathbf{H}^
{\|}_{\textrm{EX},\nu}+\mathbf{H}''_{\textrm{eff},\nu}$ where
$\mathbf{H}''_{\textrm{eff},\nu}  =\mathbf{H}+
\mathbf{H}_{\textrm{A},\nu}+\mathbf{H}_{\textrm{EX},\nu}^{\bot}$.  We now  expand
$\mathbf{m}_{0,\nu}$ up to the first order in $\mathbf{H}''_{\textrm{eff},\nu}$, under the assumption
 $\left|\mathbf{H}_{\textrm{EX},\nu}^{\Vert}\right|\gg|\mathbf{H}''_{\textrm{eff},\nu}|$.
Similar to the ferromagnetic case, from Eq. (\ref{eq:m0xi0}) we get (see Appendix A)

%==========================================================================================
\begin{eqnarray}
\mathbf{m}_{0,\nu}&\simeq&\frac{B_{\nu}}{m_{\nu}}\mathbf{m}_{\nu}
+B_{\nu}'\beta\mu_{\nu}\frac{\left(\mathbf{m}_{\nu}\cdot\mathbf{H}''_{\textrm{eff},\nu}\right)
\mathbf{m}_{\nu}}{m^{2}_{\nu}} \nonumber \\
&-&\frac{B_{\nu}\mu_{\nu}}{m_{\nu}\widetilde{J}_{0,\nu}}
\frac{\left[[\mathbf{H}''_{\textrm{eff},\nu}\times\mathbf{m}_{\nu}]\times\mathbf{m}_{\nu}\right]}{m^{2}_{\nu}},
\label{eq:NewM04-1}
\end{eqnarray}
%========================================================================================
substituting this into Eq. (\ref{eq:LLBequationFerrimagnet}) and repeating
the same calculations as in the ferromagnetic case we get the following equation of motion
%========================================================================================
\begin{eqnarray}
\mathbf{\dot{m}}_{\nu} & = & \gamma_{\nu}[\mathbf{m}_{\nu}\times\mathbf{H}''_{\textrm{eff},\nu}]\nonumber \\
& -&
\gamma_{\nu}\alpha^{\nu}_{\Vert}\left(\frac{1-B_{\nu}/m_{\nu}}{\mu_{\nu}\beta B_{\nu}'}-\frac{\mathbf{m}_{\nu}\cdot\mathbf{H}''_{\textrm{eff},\nu}}{m_{\nu}^{2}}\right)\mathbf{m}_{\nu}\nonumber \\
 &-&  \gamma_{\nu}\alpha^{\nu}_{\bot}\frac{\left[\mathbf{m}_{\nu}\times\left[\mathbf{m}_{\nu}
 \times\mathbf{H}''_{\textrm{eff},\nu}\right]\right]}{m_{\nu}^{2}}
 \label{eq:aproxLLB}
\end{eqnarray}
%=====================================================================================
where $B_{\nu}=B_{\nu}\left(\beta\widetilde{J}_{0,\nu}\left(\mathbf{m}_{\nu},\mathbf{m}_{\kappa}\right)m_{\nu}\right)$
depends on the sublattice magnetizations $\left(\mathbf{m}_{\nu},\mathbf{m}_{\kappa}\right)$ and the
damping parameters are:
%========================================================================================0
\begin{equation}
\alpha_{\Vert}^{\nu}  = \frac{2\lambda_{\nu}} { \beta \widetilde{J}_{0,\nu}},\quad \quad
\alpha_{\bot}^{\nu}  =  \lambda_{\nu}\left(1-\frac{1}{\beta \widetilde{J}_{0,\nu}}\right).
\label{eq:FerriDampingParameters}
\end{equation}
%=============================================================================================

\subsection{Temperature dependence of damping parameters}
\label{sec:Temperaturedependenceofdampingparameters}
\begin{figure}[h!]
\centering
\includegraphics[scale=1,angle=0]{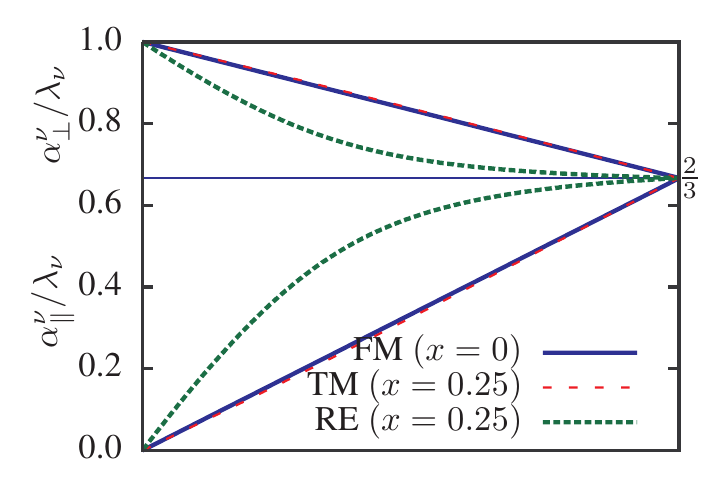}
\includegraphics[scale=1,angle=0]{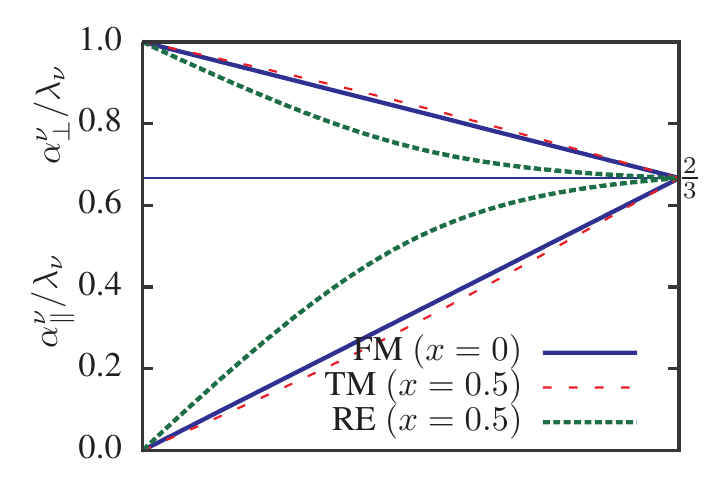}
\includegraphics[scale=1,angle=0]{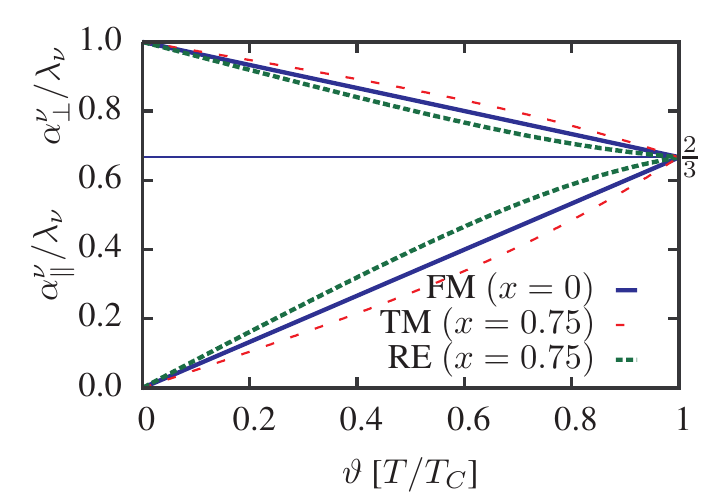}
\caption{Damping parameters $\alpha^{\nu}_{\| (\bot)} (\vartheta)$ (normalized to the corresponding intrinsic values)
for a pure ferromagnet (FM), rare earth (RE) component in a GdFeCo ferrrimagnet
and a transition metal (TM) in a ferrimagnet
as a function of reduced temperature
$\vartheta=T/T_C$ for three different rare earth (RE) concentrations $x$. The blue solid line represents the  $x=0$ limit  which corresponds to a pure ferromagnet (FM). (Up) The corresponding curves
for a $25\%$ concentration of RE.
 (Middle) The corresponding damping parameters for a $50\%$ alloy.
(Bottom) Damping values for  $75\%$  RE amount. It can be also seen as a
RE doped with a $25\%$ of transition metal (TM).}
\label{fig:ferrimagnetdamping}
\end{figure}

 The temperature dependence of the damping parameters is obtained in the first order in deviations
of magnetization  from their equilibrium value.
 Note that in Eq. \eqref{eq:aproxLLB} all terms are of the first order in the parameter
 $H''_{\text{eff},\nu}/H_{\tiem{EX},\nu}$ so that the damping parameters should be evaluated in
the zero order in this parameter.
Consequently, we can use the following equilibrium expression:

%=========================================================================================
\begin{equation}
\widetilde{J}_{0,\nu}\simeq\frac{J_{0,\nu} m_{e,\nu}+ |J_{0,\nu\kappa}|m_{e,\kappa}}{m_{e,\nu}}
\end{equation}
%=========================================================================================0
where the sign of the second term does not depend on the sign of the interlattice exchange interaction, $J_{0,\nu\kappa}$.
The effective damping parameters  depend on temperature $T$ via temperature-dependent equilibrium magnetization.
The temperature dependence of damping parameters \eqref{eq:FerriDampingParameters}, normalized to the intrinsic coupling parameter, are presented in
Fig. \ref{fig:ferrimagnetdamping} for a GdFeCo RE-TM ferrimagnet and for various concentrations of RE impurities.

 Let us consider some limiting cases.
First we consider the simplest case of a completely symmetric antiferromagnet (AFM).
In the AFM all the relevant parameters are equal for both lattices,
they have the same magnetic moments $\mu_1=\mu_2$ and the same intra-lattice exchange parameters $J_{0,\nu}$,
the inter-lattice exchange parameter is also the same $J_{0,\nu\kappa}=J_{0,\kappa\nu}$
in contrast to our disordered ferrimagnet.
In this case the equilibrium magnetizations as a function of temperature
are the same $m_{e,\nu}(T)=m_{e,\kappa}(T)$
and the effective exchange parameter reduces to $\widetilde{J}_{0,\nu}=J_{0,\nu}+|J_{0,\nu\kappa}|$,
\emph{i.e.} the sum of the two interactions coming from the intra-lattice and inter-lattice exchange.
The N\'{e}el temperature in the MFA reads $k_B T_N=\widetilde{J}_{0,\nu}/3$
and the damping parameters recover the ferromagnetic type expression

%========================================================================================0
\begin{equation}
\alpha_{\Vert}^{\nu(\tiem{AFM})}  = \lambda_{\nu}\frac{2T} { 3T_N},\quad \quad
\alpha_{\bot}^{\nu(\tiem{AFM})}  =  \lambda_{\nu}\left(1-\frac{T}{3T_N}\right).
\label{eq:AFMDampingParameters}
\end{equation}
%=============================================================================================
The use of the critical temperature provides an expression in which
the damping parameters do not depend explicitly  on the interlattice exchange,
the implicit dependence comes from the change of the N\'{e}el temperature as the exchange parameter
$J_{0,\nu\kappa}$ varies.
There is a more simple AFM, with nearest neighbor interactions only and one  inter-lattice exchange parameter
$J_{0,\nu\kappa}$, it gives  the same result as above and exactly  the same as for the ferromagnet.

Next interesting case is when one of the three exchange parameters can be neglected.
We can consider, for example, a negligible exchange between the rare-earth magnetic moments,
it is a good approximation if the impurity content is low.
Then we can write the effective exchange as
%=========================================================================================
\begin{eqnarray}
\widetilde{J}_{0,\tiem{TM}}&=&\frac{J_{0,\tiem{TM}} m_{e,\tiem{TM}}+|J_{0,\tiem{TM-RE}}|m_{e,\tiem{RE}}}{m_{e,\tiem{TM}}}\simeq J_{0,\tiem{TM}} \\
 \widetilde{J}_{0,\tiem{RE}}&=&|J_{0, \tiem{RE-TM}}|\frac{  m_{e,\tiem{TM}} }{ m_{e,\tiem{RE}} }.
\end{eqnarray}
%=================================================================================================
In this case the TM damping parameters can be approximately expressed with the antiferromagnetic or ferromagnetic  ($T_{N}\rightarrow T_{C}$)
 formula \eqref{eq:AFMDampingParameters} because in the limit $x\rightarrow 0$ the Curie temperature
of the disordered ferrimagnet is close to $k_B T_C= J_{0,\tiem{TM}} /3$ \cite{Ostler}.
The damping parameter for the the  RE lattice, however, is different. It strongly depends
on the polarization effect of the TM lattice on the RE magnetization. In this case close to $T_C$ the polarization effect can be expressed
 using the expansion, $B\approx \xi/3$, which
for this case reads
$m_{e,\tiem{RE}}\approx \beta J_{0,\tiem{RE-TM}} m_{e,\tiem{TM}}$, thus, $\widetilde{J}_{0,\tiem{RE}} \approx 1/(3\beta)$.
Therefore,  we have the following expressions
 %========================================================================================0
\begin{eqnarray}
\alpha_{\Vert}^{\tiem{TM}} & =& \lambda_{\tiem{TM}}\frac{2T} { 3T_C},\quad \quad
\alpha_{\Vert}^{\tiem{RE}} =\frac{2}{ 3} \lambda_{\tiem{RE}}. \\
\alpha_{\bot}^{\tiem{TM}}  &=&  \lambda_{\tiem{TM}}\left(1-\frac{T}{3T_C}\right), \quad \quad
\alpha_{\bot}^{\tiem{RE}}  =  \frac{2}{3} \lambda_{\tiem{RE}}.
\label{eq:AFMDampingParametersCloseTc}
\end{eqnarray}
%=============================================================================================
This relation becomes quite important above  $T_C$. We observe in Fig. \ref{fig:ferrimagnetdamping}
that even for quite large amounts of RE of $25\%$ and $50\%$, the above approximation holds quite well.

%====================== Sketch of a ferrimagnet============
%=========================================================================================

If the inter-lattice exchange is  large in comparison to the intra-lattice one then
the equilibrium magnetization of both lattices is similar and the damping parameters
 behave similar to those of the FM damping parameters, presented above. This case is in
agreement with a concentration of $75\%$ of RE in Fig. \ref{fig:ferrimagnetdamping} (down).
As predicted, we observe that the damping parameters are very similar for both sublattices.

Note that these damping parameters should be distinguished from those of the normal modes
(FMR and exchange) with more complicated expressions which can be obtained via linearization
of the set of two-coupled LLB equations \cite{Schlickeiser}, similar to the LLG approach.

\subsection{Longitudinal relaxation parameters}

 The function $1-B_{\nu}/m_{\nu}$ in Eq. (\ref{eq:aproxLLB})
is a small quantity proportional to the deviation from the equilibrium
in  both sublattices.
 It can be further simplified as a function of the equilibrium parameters after some algebra.
Similar to the ferromagnetic case, the ferrimagnetic LLB equation can be put in a compact form using the notion of the longitudinal susceptibility.

 The initial longitudinal susceptibility can be evaluated on the basis of the
Curie-Weiss equations (\ref{eq:CurieWeissImp}). Let us assume that in the absence of an
external field,  the equilibrium sublattice magnetizations $\mathbf{m}_{\tiem{TM}}$
and $\mathbf{m}_{\tiem{RE}}$ are, respectively,  parallel and antiparallel
to the $z$-axis (a stronger condition of the smallness of the perpendicular components can be also applied).
The $z$-axis is chosen such that it is the easy
axis of the magnetic crystal. To evaluate the longitudinal susceptibility,
the field should be applied parallel to the easy direction,
then  in the  approximation of small perpendicular components (large longitudinal exchange field) we can neglect in the first approximation the possible change of directions of  $\mathbf{m}_{\tiem{RE}}$
and $\mathbf{m}_{\tiem{TM}}$.
In order to calculate the susceptibility, we expand the right-hand
side of Eq. (\ref{eq:CurieWeissImp}) in terms of the external field:

%============================================================================================
\begin{equation}
m_{\nu}(T,H_{z})\approx
m_{\nu}(T,0)+\mu_{\nu}H_{z}\beta B'_{\nu}\left(1+\frac{\partial H_{\textrm{EX},\nu}^{z}}{\partial H_{z}}\right),
\end{equation}
%===========================================================================================
where $B_{\nu}=B_{\nu} (\beta \mu_{\nu} H_{\tiem{EX},\nu})$ and its derivative $B'_{\nu}=B'_{\nu} (\beta \mu_{\nu} H_{\tiem{EX},\nu})$
 are evaluated in absence of applied  and anisotropy fields.
Then,
%============================================================================================
\begin{equation}
\widetilde{\chi}_{\nu,||}=\left(\frac{\partial m_{\nu}(T,H_{z})}{\partial H_{z}}\right)_{H_{z}=0}
=\mu_{\nu}\beta B'_{\nu}\left(1+\frac{\partial H_{\textrm{EX},\nu}^{z}}{\partial H_{z}}\right),
\end{equation}
%================
where
%===========================================================================================
\begin{eqnarray*}
\frac{\partial H_{\textrm{EX},\nu}^{z}}{\partial H_{z}}=\beta J_{0,\nu} \widetilde{\chi}_{\nu,||} +
\beta |J_{0,\nu\kappa}|\widetilde{\chi}_{\kappa,||}.
 \end{eqnarray*}
%==========================================================================================0
 Thus, the longitudinal susceptibility of one sublattice is expressed in terms of another:

%===========================================================================================
\begin{eqnarray}
\widetilde{\chi}_{\nu,||} & = &
\frac{\mu_{\nu}}{J_{0,\nu}}\frac{J_{0,\nu}\beta B'_{\nu}}{1-J_{0,\nu}\beta B'_{\nu}}
\left[\frac{|J_{0,\nu \kappa}|}{\mu_{\nu}}\widetilde{\chi}_{\kappa,||}+1\right].
 \label{eq:longitudinalSusc1}
 \end{eqnarray}
%==========================================================================================0
Finally, we obtain two coupled equations for $\widetilde{\chi}_{\tiem{RE},||}$ and $\widetilde{\chi}_{\tiem{TM},||}$,
solving them, we get the MFA expression for the susceptibilities:
\begin{widetext}
%=============================================================================================
\begin{equation}
\widetilde{\chi}_{\nu,||}  =  \left( \frac{\mu_{\kappa}}{|J_{0,\kappa\nu}|}\right)
 \frac{|J_{0,\kappa\nu}|\beta B'_{\nu}|J_{0,\nu\kappa}|\beta B'_{\kappa}+(\mu_{\nu}/\mu_{\kappa})|J_{0,\kappa\nu}|
\beta B'_{\nu}\left(1-J_{0,\kappa}\beta B'_{\kappa}\right)}
 {\left(1-J_{0,\nu}\beta B'_{\nu}\right)\left(1-J_{0,\kappa}\beta B'_{\kappa}\right)-
\left(|J_{0,\kappa\nu}|\beta B'_{\nu}\right)\left(|J_{0,\nu\kappa}|\beta B'_{\kappa}\right)}=
\left( \frac{\mu_{\kappa}}{|J_{0,\kappa\nu}|}\right) G_{\nu} (T)
 \label{eq:generalSusceptibiltyFerrim}
 \end{equation}
%=====================================================================================
\end{widetext}
The longitudinal susceptibility $\widetilde{\chi}_{\nu,||}$ is, therefore, a function of temperature
which we have called $G_{\nu}(T)$. It tends to zero at low temperature and
diverges approaching Curie temperature $T_C$ of the magnetic system, similar to the ferromagnetic case.
 The function $G_{\nu}=(|J_{0,\nu\kappa}|/\mu_{\nu})\widetilde{\chi}_{\nu,||}$
can be seen as a reduced longitudinal susceptibility.

Now we derive  an approximate expression  for the small quantity $1-B_{\nu}/m_{\nu}$
as a function of equilibrium quantities and the deviation of each sublattice magnetization from its equilibrium.
In the first approximation,  we  expand the function $B_{\nu}/m_{\nu}$ near the equilibrium,
as was done for the ferromagnet.
The function $B_{\nu}$ in the zero order in perpendicular field components, $H''_{eff,\nu}/H_{\tiem{EX},\nu}$, can be written
as a function of $m_{\nu}$ and $m_{\kappa}$ as follows

%===========================================================================
\begin{equation}
B_{\nu} \approx B_{\nu}\left(\beta [J_{0,\nu}m_{\nu} + |J_{0,\nu\kappa}|\tau_{\kappa} ]\right)
 \end{equation}
%==================================================================================
 where $\tau_{\kappa}=|(\mathbf{m}_{\nu}\cdot\mathbf{m}_{\kappa})|/m_{\nu}$
is the length of the projection  of the magnetization of the sublattice $\kappa$
onto the
sublattice $\nu$. We expand the function $B_{\nu}/m_{\nu}$ in the
variables $m_{\nu}$ and $m_{\kappa}$ near the equilibrium :

%========================================================================
\begin{eqnarray}
\frac{B_{\nu}}{m_{\nu}} & \approx & \frac{B_{e,\nu}}{m_{e,\nu}}+
\left[\frac{1}{m_{\nu}}\left(\frac{\partial B_{\nu}}{\partial m_{\nu}}\right)
-\frac{1}{m_{\nu}^{2}}B_{\nu}\right]_{\mathrm{eq}}
\delta m_{\nu} \\\nonumber &+&
\left[\frac{1}{m_{\nu}}\frac{\partial B_{\nu}}{\partial\tau_{\kappa}}\right]_{\text{eq}}\delta\tau_{\kappa}\\\nonumber
 & = & 1-\left[1-\beta J_{0,\nu}B'_{\nu}\right]_{\text{eq}}\frac{\delta m_{\nu}}{m_{e,\nu}}+\left[\beta |J_{0,\nu\kappa}|B'_{\nu}\right]_{\text{eq}}
\frac{\delta\tau_{\kappa}}{m_{e,\nu}},
 \end{eqnarray}
 %==========================================================================
 here  $\delta m_{\nu}=m_{\nu}-m_{e,\nu}$,
with $m_{e,\nu}=B_{\nu} (\beta \mu_{\nu} H_{ \tiem{EX},\nu})$, where $H_{ \tiem{EX},\nu}$ is evaluated at the equilibrium,
and $\delta \tau_{\kappa}=\tau_{\kappa}-\tau_{e,\kappa}$,
 where $\tau_{e,\kappa}=|(\mathbf{m}_{e,\nu}\cdot\mathbf{m}_{e,\kappa})|/m_{e,\nu}$ and it
corresponds to the projection of the equilibrium magnetization  $\mathbf{m}_{e,\kappa}$ onto the
other sublattice magnetization direction.
 It is easy to show that
$\partial\tau_{\kappa}/\partial m_{\nu}=0$.
Similar to the ferromagnetic case, we would like to arrive to a simplified expression as a function of sublattice susceptibilities.
For this purpose, we divide the above expression by $\mu_{\nu}\beta B_{\nu}'$

%====================================================================================
\begin{eqnarray}
\frac{1-B_{\nu}/m_{\nu}}{\mu_{\nu}\beta B_{\nu}'}
&=& \frac{1}{\widetilde{\chi}_{\nu,||}} \frac{\delta m_{\nu}}
{m_{e,\nu}}+ \nonumber \\
  &+& G_{\kappa}\left[
 \frac{1}{\widetilde{\chi}_{\nu,||}}\frac{\delta m_{\nu}}
{m_{e,\nu}}
 -\frac{1}{\widetilde{\chi}_{\kappa,||} } \frac{\delta\tau_{\kappa}}{m_{e,\nu}} \right]
\label{eq:exchangelongitudinalcontribution}
\end{eqnarray}
%=================================================================================
where we have used  Eq. (\ref{eq:longitudinalSusc1})
and the function
$G_{\kappa}=|J_{0,\nu\kappa}|\widetilde{\chi}_{\kappa,||}/\mu_{\nu}$ has now more sense.
Thus,
the contribution to the  dynamical equation \eqref{eq:LLBequationFerrimagnet} of
the  exchange interaction (the LLB equation with longitudinal relaxation only)  given by
Eq. \eqref{eq:exchangelongitudinalcontribution} reads

%========================================================================================
\begin{eqnarray}
\frac{\dot{\mathbf{m}}_{\nu}}{\gamma_{\nu}} \rvert_{EX}
  =  -\frac{\alpha_{\Vert}^{\nu}}{m_{e,\nu}}
\left(
\frac{1+G_{\kappa}}{\widetilde{\chi}_{\nu,||}}
 \delta m_{\nu} -
\frac{|J_{0,\nu\kappa}|}{\mu_{\nu}}
\delta \tau_{\kappa}
\right)\mathbf{m}_{\nu}
 \label{eq:ferrilongitudinalmotionEq}
\end{eqnarray}
%=====================================================================================
Note that the first term defines the intralattice relaxation of the sub-lattice (for example, TM) to its own
direction. The second term describes the angular momenta transfer between sublattices driven by the temperature.
This equation has the form
\begin{equation}
\frac{\dot{\mathbf{m}}_{\nu}}{\gamma_{\nu}}=\widetilde{\Gamma}_{\nu}
 \mathbf{m}_{\nu}
 \end{equation}
and it  gives the exact
 LLB equation for the case when the average magnetization of the two sublattices remain always antiparallel.

\subsection{Final forms of the LLB equation}

In order to be consistent with the ferromagnetic LLB equation (and the Landau theory of phase  transitions), we
expand the deviations $\delta m_{\nu}$ ($\delta \tau_{\kappa}$) around  $m_{e,\nu}^2 $ ($\tau_{e,\nu}^2$) up to the quadratic terms. Similar to FM case we write:

%=========================================================================================
\begin{equation}
\frac{\delta m_{\nu}}{m_{\nu,e}}\approx \frac{1}{2m_{e,\nu}^{2}}\left(m_{\nu}^{2}-m_{e,\nu}^{2}\right)
%,\quad
%\frac{\delta \tau_{\kappa}}{\tau_{\kappa,e}}\approx \frac{1}{2\tau_{e,\kappa}^{2}}\left(\tau_{\kappa}^{2}-\tau_{e,\kappa}^{2}\right)
\label{Landau}
\end{equation}
%===========================================================================================
Therefore we can write the effective longitudinal fields as

%========================================================================================
\begin{equation}
\mathbf{H}_{\textrm{eff},||}^{\nu}=
\left[\frac{1}{2\Lambda_{\nu\nu}}\left(\frac{m_{\nu}^{2}}{m_{e,\nu}^{2}}-1\right)
-\frac{1}{2\Lambda_{\nu\kappa}}\left(\frac{\tau_{\kappa}^2}{\tau_{e,\kappa}^2}-1\right)\right]\mathbf{m}_{\nu}
\end{equation}
%=============================================================================================
where in order to shorten the notations we have defined the  longitudinal rates as:

%================================================================================================
\begin{equation}
\Lambda_{\nu\nu}^{-1}=\frac{1}{\widetilde{\chi}_{\nu,||}}\left(1+G_{\kappa}\right),
\quad\Lambda_{\nu\kappa}^{-1}=
\frac{\tau_{e,\kappa}}{m_{e,\nu}}\frac{|J_{0, \nu \kappa}|}{\mu_{\nu}} \textrm{ with }\nu\neq\kappa,
\label{eq:longitudinalrelaxparametersLLBferrimag}
\end{equation}
%===========================================================================================
where $G_{\kappa}$ is also expressed in terms of the longitudinal susceptibility via Eq.(\ref{eq:generalSusceptibiltyFerrim}).

\subsubsection*{Form 1}
Finally, we collect all the above derived approximate expressions and we finish up with the compact form of the LLB equation
for the reduced  magnetization vector, $\mathbf{m}_{\nu}=\mathbf{M}_{\nu}/M_{\nu}(T=0K)$
%=================================================================================================================0
\begin{eqnarray}
\mathbf{\dot{m}}_{\nu}  &=&  \gamma_{\nu} [\mathbf{m}_{\nu}\times\mathbf{H}_{\textrm{eff},\nu}]-
\gamma_{\nu} \alpha_{\Vert}^{\nu}\frac{\left(\mathbf{m}_{\nu}\cdot\mathbf{H}_{\textrm{eff},\nu}\right)}{m_{\nu}^{2}}\mathbf{m}_{\nu} \nonumber \\
  &-&  \gamma_{\nu} \alpha_{\bot}^{\nu}\frac{\left[\mathbf{m}_{\nu}\times\left[\mathbf{m}_{\nu}\times\mathbf{H}_{\textrm{eff},\nu}\right]\right]} {m_{\nu}^{2}}
  \label{eq:LLBferrimagFinalexpression1}
 \end{eqnarray}
%=====================================================================================================
where the effective field $\mathbf{H}_{\textrm{eff},\nu}$ for sublattice
$\nu$ is defined as

%=======================================================================================================0
\begin{eqnarray}
&\mathbf{H}&{}_{\textrm{eff},\nu}  =  \mathbf{H}+\mathbf{H}_{\textrm{A},\nu}+\frac{J_{0,\nu\kappa}}{\mu_{\nu}}\boldsymbol{\Pi}_{\kappa} \nonumber \\
&+&
\left[\frac{1}{2\Lambda_{\nu\nu}}\left(\frac{m_{\nu}^{2}}{m_{e,\nu}^{2}}-1\right)-
\frac{1}{2\Lambda_{\nu\kappa}}\left(\frac{\tau_{\kappa}^2}{\tau_{e,\kappa}^2}-1\right)\right]\mathbf{m}_{\nu}
 \end{eqnarray}
%=======================================================================================================
and the relaxation parameters $\alpha_{\|}^{\nu}$ and $\alpha_{\bot}^{\nu}$ are given by Eqs. \eqref{eq:FerriDampingParameters}.

%\subsubsection*{Form 2}
Or in a more explicit form, as a function of sub-lattice magnetizations
 $\mathbf{m}_{\nu}$ and its values at the equilibrium $\mathbf{m}_{e,\nu}$:
%=================================================================================================================0
\begin{eqnarray}
\mathbf{\dot{m}}_{\nu} & =&  \gamma_{\nu} [\mathbf{m}_{\nu}\times\mathbf{H}_{\textrm{eff},\nu}]-
\gamma_{\nu} \alpha_{\Vert}^{\nu}\frac{\left(\mathbf{m}_{\nu}\cdot\mathbf{H}^ {\|}_{\textrm{eff},\nu}\right)}{m_{\nu}^{2}}\mathbf{m}_{\nu}\nonumber \\
 & -&  \gamma_{\nu} \alpha_{\bot}^{\nu}\frac{\left[\mathbf{m}_{\nu}
\times\left[\mathbf{m}_{\nu}\times\mathbf{H}_{\textrm{eff},\nu}\right]\right]} {m_{\nu}^{2}}
  \label{eq:LLBferrForm2}
 \end{eqnarray}
%=====================================================================================================
where we have defined
the
longitudinal field, $\mathbf{H}^{\|}_{\textrm{eff},\nu}$,  as

 %=======================================================================================================0
\begin{eqnarray}
\mathbf{H}^{\|}_{\textrm{eff},\nu}&=&
\Big[\frac{1}{2\Lambda_{\nu\nu}}\left(\frac{m_{\nu}^{2}}{m_{e,\nu}^{2}}-1\right)
 \nonumber\\
&-&\frac{1}{2\Lambda_{\nu\kappa}}
\left(
\left( \frac{ \mathbf{m}_{\nu} \cdot \mathbf{m}_{\kappa} }
{\mathbf{m}_{e,\nu} \cdot \mathbf{m}_{e,\kappa}}\right)^2
-1\right)\Big]\mathbf{m}_{\nu}
\label{form1}
 \end{eqnarray}
%=======================================================================================================
and the effective field, $\mathbf{H}_{\text{eff},\nu}$, reads

%=======================================================================================================0
\begin{eqnarray*}
\mathbf{H}_{\textrm{eff},\nu} & = & \mathbf{H}+\mathbf{H}_{\textrm{A},\nu}+
\frac{J_{0,\nu\kappa}}{\mu_{\nu}}\mathbf{m}_{\kappa}.
 \end{eqnarray*}
%===================
In Eq. \eqref{eq:LLBferrForm2} also the temperature dependent damping parameters are given by Eqs. \eqref{eq:FerriDampingParameters}.
 \subsubsection*{Form 2}
It is also interesting to put the LLB equation in a more symmetric form in terms of the macroscopic magnetization,
$\mathbf{M}_{\nu}=x_{\nu}\mu_{\nu} \mathbf{m}_{\nu}/\upsilon_{\nu}$, where $x_{\nu}$ stands for the concentration of
sites of type $\nu=$TM or RE ($x_{\nu}=x$ for RE and $x_{\nu}=q$ for TM), $\mu_{\nu}$ is the atomic magnetic moment of the lattice  $\nu$ and
$\upsilon_{\nu}$ is the atomic volume.
We multiply each sublattice LLB equation (\ref{eq:LLBferrForm2}) by the corresponding factor, for example,
in the case of TM by $q\mu_{\tiem{TM}}/\upsilon_{\tiem{TM}}$ and we obtain
%=================================================================================================================0
\begin{eqnarray}
\mathbf{\dot{M}}_{\nu} & =&  \gamma_{\nu} [\mathbf{M}_{\nu}\times\mathbf{H}_{\textrm{eff},\nu}]-
L_{\|,\nu}\frac{\left(\mathbf{M}_{\nu}\cdot\mathbf{H}^ {\|}_{\textrm{eff},\nu}\right)}{M_{\nu}^{2}}\mathbf{M}_{\nu}\nonumber \\
  &-& L_{\bot,\nu}\frac{\left[\mathbf{M}_{\nu}
\times\left[\mathbf{M}_{\nu}\times\mathbf{H}_{\textrm{eff},\nu}\right]\right]} {M_{\nu}^{2}}
 \end{eqnarray}
%=====================================================================================================
where the effective fields read:

 %=======================================================================================================0
\begin{eqnarray}
\mathbf{H}^{\|}_{\textrm{eff},\nu} & = &
\Big[\frac{1}{2\widetilde{\Lambda}_{\nu\nu}}\left(\frac{M_{\nu}^{2}}{M_{e,\nu}^{2}}-1\right)\nonumber \\
&-&
\frac{1}{2\widetilde{\Lambda}_{\nu\kappa}}
\left(
\left( \frac{ \mathbf{M}_{\nu} \cdot \mathbf{M}_{\kappa} }
{\mathbf{M}_{e,\nu} \cdot \mathbf{M}_{e,\kappa}}\right)^2
-1\right)\Big]\mathbf{M}_{\nu},
 \end{eqnarray}
%=======================================================================================================
The rate parameters are $\widetilde{\Lambda}_{\nu\kappa}=\upsilon_{\nu}\Lambda_{\nu\kappa}/\mu_{\nu}x_{\nu}$
and the effective field, $\mathbf{H}_{\text{eff},\nu}$, has the following form:

  %=======================================================================================================0
\begin{eqnarray*}
\mathbf{H}_{\textrm{eff},\nu} & = & \mathbf{H}+\mathbf{H}_{\textrm{A},\nu}+
A\mathbf{M}_{\kappa}.
 \end{eqnarray*}
%===================
Here the exchange parameter is introduced as $A=zJ_{\tiem{TM-RE}}/\mu_{\tiem{RE}}\mu_{\tiem{TM}}$.
The damping coefficients $L_{\|,\nu}$
and $L_{\bot,\nu}$ read

%==============================================================================
\begin{eqnarray*}
L_{\|,\nu}=\gamma_{\nu} x_{\nu}\mu_{\nu} \alpha_{\|}^{\nu}/\upsilon_{\nu}, \quad
L_{\bot,\nu}=\gamma_{\nu} x_{\nu}\mu_{\nu} \alpha_{\bot}^{\nu}/\upsilon_{\nu}.
 \end{eqnarray*}
%============================================================================

\section{Relaxation of magnetic sublattices}

The rate of the longitudinal relaxation
is  temperature dependent through the parameters such as the  damping parameters $\alpha_{\Vert}^{\nu}$,
see Eq. \eqref{eq:FerriDampingParameters} and Fig. \ref{fig:ferrimagnetdamping}, and the longitudinal susceptibilities.
The  sign of the rate,  $\widetilde{\Gamma}_{\nu}\lessgtr 0$,  depends  on the instantaneous magnetization values.
 From Eq. \eqref{eq:ferrilongitudinalmotionEq} we can consider the following lines separating different relaxation signs:
%=====================================================================================================
\begin{equation}
\delta m_{\nu}=
\frac{|J_{0,\nu\kappa}|}{\mu_{\nu}}\frac{\widetilde{\chi}_{\nu,||}}
{G_{\kappa}+1}
\delta\tau_{\kappa}
=\widetilde{\chi}_{\nu\kappa,||}\delta\tau_{\kappa},
\label{eq:susceptibility12relation}
\end{equation}
%=====================================================================================================0
where we have defined the dimensionless variable $\widetilde{\chi}_{\nu\kappa,||}$,
which describes the effect of the change in one sublattice on the other.
This variable can be interpreted as a susceptibility $\widetilde{\chi}_{\nu\kappa,||}  = \delta m_{\nu}/\delta m_{\kappa}$.
 Indeed, we can expand:
%==================================================================================================00
\begin{eqnarray}
m_{\nu}(T,\delta m_{\nu},\delta m_{\kappa})&\approx& m_{\nu}(T,0,0)+\beta J_{0,\nu}
B'_{\nu}\delta m_{\nu} \nonumber \\
%& &+
& &+\beta |J_{0,\nu\kappa}|B'_{\nu}\delta m_{\kappa}
%+\mathcal{O}
%\left(\left(\delta m_{\nu}\right)^{2},\left(\delta m_{\kappa}\right)^{2}\right)
\end{eqnarray}
%===================================================================================================
Now using that by definition $\delta m_{\nu}=m_{\nu}(T,\delta m_{\nu},\delta m_{\kappa})-m_{\nu}(T,0,0)$,
we obtain
%====================================================================================================0
\begin{eqnarray}
\widetilde{\chi}_{\nu\kappa,||}=|J_{0,\nu\kappa}|\left(\frac{\beta
B'_{\nu}}{1-J_{0,\nu}
\beta B'_{\nu}}\right)
\label{eq:longitudinalSusc21}
\end{eqnarray}
%=======================================================================================0
Next, we substitute Eq. \eqref{eq:longitudinalSusc1}
into Eq. \eqref{eq:longitudinalSusc21}
and we get the  relation between the susceptibilities, exactly described by Eq. \eqref{eq:susceptibility12relation}.

\begin{figure}[h!]
\centering
\includegraphics[scale=0.9,angle=90]{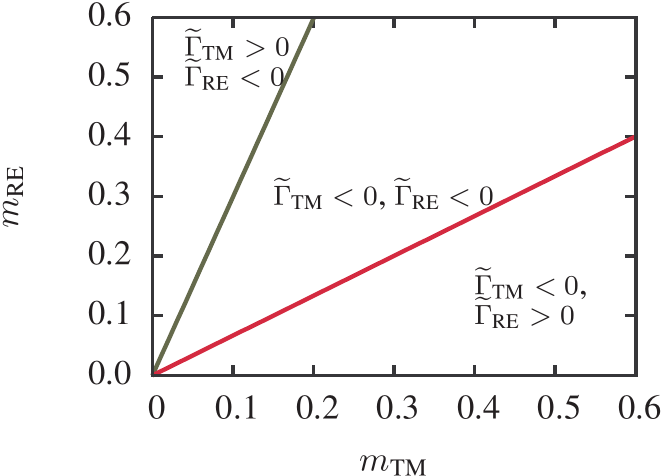}
\caption{Different longitudinal relaxation regions for $T/T_C=0.95$ for parameters of the GdFeCo alloy with $x=0.25$.  }
\label{fig:phasespace}
\end{figure}

The problem of relaxation sign is, therefore,  reduced  to the study of the sign of the function
$\delta m_{\nu} -\widetilde{\chi}_{\nu\kappa,||}\delta \tau_{\kappa}$.
Let us assume the equilibrium state that is close  to $T_C$, describing the situation during the ultrafast laser-induced
demagnetization \cite{RaduNATURE2011}.
Fig. \ref{fig:phasespace} shows  three possible
instantaneous rates for $T= 0.95 T_C$, depending on the relative state of both sublattice magnetizations.
 The lines separating different relaxation types  are  straight lines with the slope
 $\widetilde{\chi}_{\nu\kappa,||}(T)$.

\begin{figure}[h!]
\centering
\includegraphics[scale=0.6,angle=0]{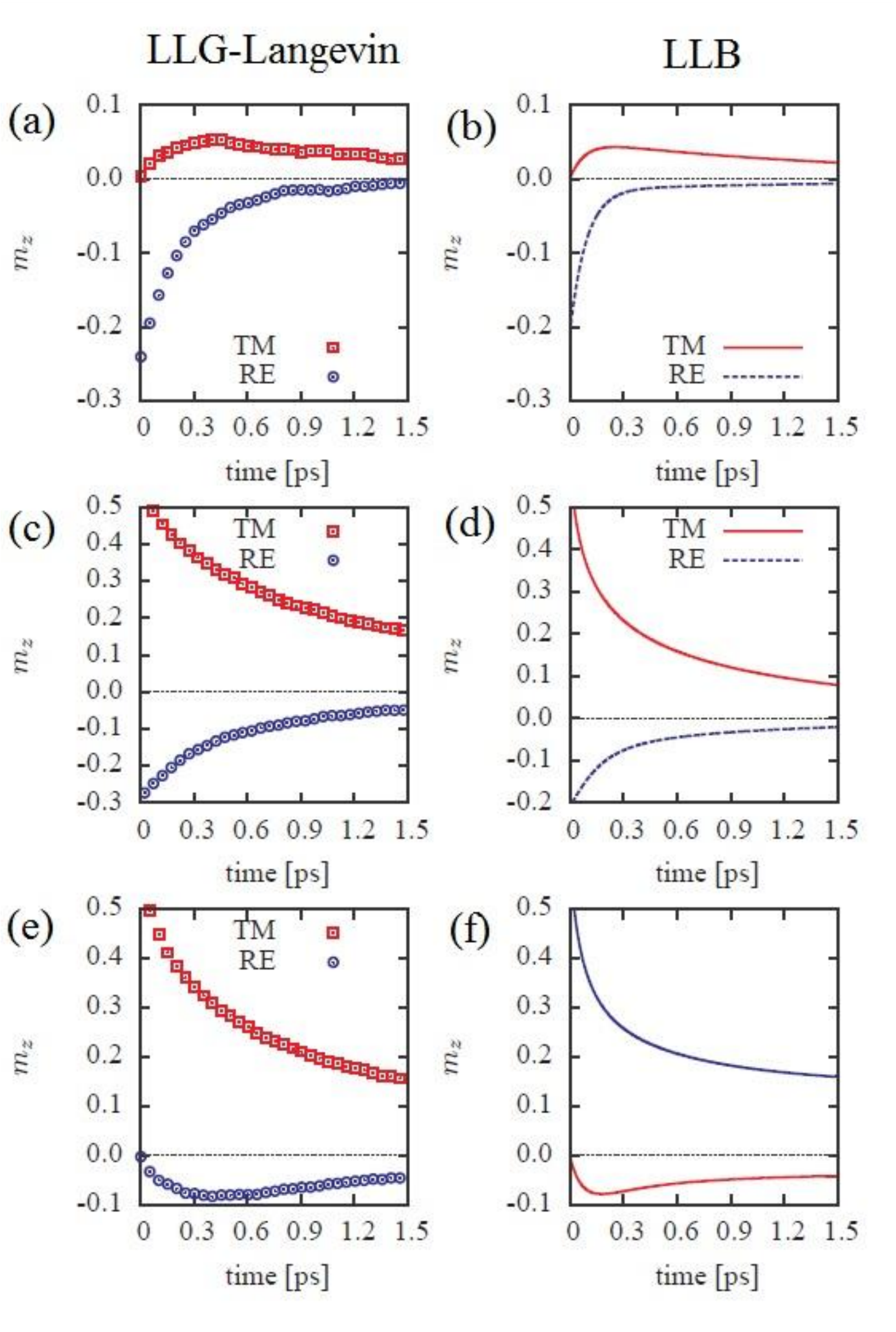}
\caption{ Comparison between atomistic LLG-Langevin and macrospin LLB calculations of the longitudinal relaxation of the GdFeCo alloy ($x=0.25$) corresponding to the
three different relaxation cases in Fig.\ref{fig:phasespace}.
In the left column we show atomistic LLG-Langevin multispin simulations and in the right one- the LLB macrospin calculations.
The graphs (a) and (b) correspond to the region with $\widetilde{\Gamma}_{\tiem{TM}}>0$ and $\widetilde{\Gamma}_{\tiem{RE}}<0$.
The graphs (c) and (d) correspond to the region with $\widetilde{\Gamma}_{\tiem{TM}}<0$ and $\widetilde{\Gamma}_{\tiem{RE}}<0$.
The graphs (e) and (f)  correspond to the region with $\widetilde{\Gamma}_{\tiem{TM}}<0$ and $\widetilde{\Gamma}_{\tiem{RE}}>0$.  }
\label{fig:3cases}
\end{figure}

In the following we use atomistic LLG Langevin simulations described in Sec. \ref{sec:AtomisticModelFerri}
 as well as the integration of the LLB equation \eqref{eq:LLBequationFerrimagnet}
for the  same material parameters, see Table \ref{table_parameters}. In order to compare MFA based LLB equation and the atomistic simulations,  we have  re-normalized exchange parameters,
as described in Ref. \cite{OstlerPRB2011}.
In the atomistic simulation the system size is taken as $\mathcal{N}=60^3$, \emph{i.e.}   $3\mathcal{N}$ coupled differential equations
has to be solved simultaneously within this approach, whereas only 6 (two sublattices and three components for each) in the macrospin LLB approach.
We compare the different relaxation regions depending on the instantaneous magnetic state with those predicted
by the LLB equation and depicted in Fig. \ref{fig:phasespace}.
The initial conditions in the simulations are the following: in all three cases we start from a
an equilibrium state at $T=600$ K (for the considered concentration $x=0.25$ we get $T_C=800$ K). After that for the situations of Fig.\ref{fig:3cases}(a) and (e)  we
put one of the  sublattice magnetizations  equal to zero, $m_{\tiem{TM(RE)}}=0$. In the atomistic approach this is done by totally
disordering the system.
Finally, the temperature is set to $T=0.95T_C$ and the relaxation of both sublattices is visualized. The results are presented in Fig. \ref{fig:3cases}.

For the region $m_{\tiem{RE}}\gg m_{\tiem{TM}}$ above the green line in Fig.\ref{fig:phasespace} the rate for the TM
is positive, $\widetilde{\Gamma}_{\tiem{TM}}>0$, thus the
TM magnetization will increase while $\widetilde{\Gamma}_{\tiem{RE}}<0$ and the RE magnetization will decrease. Thus, we have initially a dynamical polarization of TM by RE. As it can be seen in
Fig.\ref{fig:3cases}(a),(b) initially the TM magnetic order increases from a totally disordered state, while the
RE relaxes directly to the equilibrium, \emph{i.e.} the sign of the RE rate is always the same.
In the central region of Fig.\ref{fig:phasespace}, between green and red lines, both magnetizations go to the equilibrium by decreasing their value,
see Fig. \ref{fig:3cases}(c)(d).
Finally, in the low region of Fig.\ref{fig:phasespace}  the situation is
symmetric to the upper region but now TM
magnetization  decreases and the RE magnetization increases initially,  see Fig. \ref{fig:3cases}(e)(f). Thus, the predictions of the LLB equation are in agreement with full atomistic simulations which also provides a validation for our analytic derivation.

%====================== F_{\nu} function============
%=========================================================================================
%====================== F_{\nu} function============
%=========================================================================================

  As a representative example, in GdFeCo near the magnetization reversal the situation is the
following \cite{RaduNATURE2011}: the TM magnetization is almost zero, $m_{\tiem{TM}}\approx0$ and the RE has finite
magnetization value $m_{\tiem{TM}}>0$. This happens due to the fact that the Gd sublattice is intrinsically slower than
the FeCo one due to a larger magnetic moment. This situation corresponds to the upper region in Fig. \ref{fig:phasespace}
 where the rates are $\widetilde{\Gamma}_{\tiem{TM}}>0$ and $\widetilde{\Gamma}_{\tiem{RE}}<0$.
Under these circumstances
 the RE  magnetization  dynamically polarizes the TM sublattice magnetization
through the interlattice exchange interaction $H_{\tiem{EX,TM-RE}}\approx |J_{0,\tiem{TM-RE}}| m_{\tiem{RE}}> 0$.
Consequently, the TM magnetization goes opposite to its equilibrium position $m_e^{\tiem{TM}}=0$ [see Fig. \ref{fig:3cases}(a)-(b)].
The existence of opposite relaxation signs in TM and RE is consistent with a recently reported ferromagnetic state
in a ferrimagnetic materials Ref. \cite{RaduNATURE2011}, however it does not necessary lead to it.
Nor it necessary means the switching of the TM magnetization, as was suggested in
Ref.\cite{MentinkPRL2012}. To have a switching one should cross the line $m_z^{\tiem{TM}}=0$
which cannot be done within the approach of longitudinal relaxation only which only describes the relaxation to the equilibrium.
The crossing of the line $m_z^{\tiem{TM}}=0$ can be only provided by a stochastic kick which is always present in
the modeling using stochastic atomistic approach \cite{Ostler, RaduNATURE2011}. This topic will be the subject of future work.

\section{The LLB equation and the Baryakhtar equation}
\label{sec:Baryaktarequation}

In this section we would like to discuss the  differences between the LLB equation and the equation
derived by V. Baryakhtar \cite{BaryakhtarZETF1988} and used in Ref. \cite{MentinkPRL2012} to explain the
ultrafast magnetization reversal and the transient ferromagnetic-like state in ferrimagnets.
 The Baryakhtar equation was derived from the Onsager principle which in general is valid near the thermodynamic equilibrium only.
The general derivation is based on the symmetry approach. Another strong supposition made in its derivation
is the separation of the timescales: the exchange interaction timescale and the relativistic interaction timescale
(defined in our case by the parameter $\lambda$) are assumed to be separated. The resulting equation has the following form:

%==============================================================================
\begin{equation}
\frac{1}{\gamma_{\nu}}\frac{d M_{\nu}}{dt}=\lambda_e \left( H_{\nu}-H_{\kappa} \right)
   +\lambda_{\nu} H_{\nu}
\label{eq:Baryaktar}
\end{equation}
%=================================================================================
Here $\nu=$ TM, RE, $\lambda_{\nu}$ describes transfer of the
angular momentum from sublattices  to the environment,
$\lambda_e$ is of the exchange origin and stems from spin-spin
interactions, conserving the total angular momentum
but allowing for the transfer of angular momentum
between the sublattices. The effective fields
defined as $H_{\nu} = - \delta W/\delta M_{\nu}$ are derived from the magnetic
energy $W$. In Ref. \cite{MentinkPRL2012} the authors used the Landau type free energy expansion
near the critical temperature, corresponding to the form Eq. (\ref{Landau}).

In comparison to the Baryakhtar equation, the LLB equation, derived here
  includes the transverse exchange mode and allows the transfer of the energy or
momentum between the longitudinal and transverse motion.
 The ferrimagnetic LLB equation has three terms among which it is  the precession term which conserves the total angular momentum.
The precession in the interlattice exchange field given by
$[\mathbf{m}_{\tiem{TM}}\times \mathbf{m}_{\tiem{RE}}]$  allows the transfer of
angular momentum between sublattices.
The longitudinal and transverse relaxation terms which are
 related to the coupling to the heat bath are both proportional to $\lambda$.
 Differently to ferromagnets, both the transverse motion given by precession and
transverse relaxation terms are not negligible on the femtosecond timescale in comparison to
 longitudinal motion because in both cases
the field acting on both motions is of the exchange origin.

In principle the ferrimagnetic LLB equation can be cast in a form, similar to the Baryakhtar
equation if we restrict ourselves to longitudinal motion only, considering the antiparallel sublattices alignment.
For the longitudinal relaxation only (see Eq. (\ref{eq:ferrilongitudinalmotionEq})) we have the following expression

%========================================================================================
\begin{eqnarray}
\frac{\dot{m}_z^{\nu}}{\gamma_{\nu}}
 =  \alpha_{\|}^{\nu} H'_{\nu}+\alpha_{\nu\kappa}^{\|}\left(H'_{\nu}
+ H'_{\kappa}\right)
\label{eq:LLBbaryaktartype}
\end{eqnarray}
%=====================================================================================
where $H'_{\nu}=-(\frac{\delta m_{\nu}}{\widetilde{\chi}_{\nu,||}}) m_z^{\nu}/m_{\nu}$, stands for the fields
coming from interaction of each lattice with itself and $H'_{\kappa}$ -with the opposite sublattice.
One can see  that the sign of
the effective field coming from the other  sublattice is opposite for the LLB Eq. \eqref{eq:LLBbaryaktartype} and
the Baryakhtar equation Eq. \eqref{eq:Baryaktar}.
In order to illustrate the consequence of this, we can compare the equations for the limiting case close to $T_C$. In this case the Baryakhtar
equation (see Eq. (1.33) in Ref. \cite{BaryakhtarZETF1988})   reads:

%========================================================================================
\begin{eqnarray}
\frac{\dot{m}_z^{\nu}}{\gamma_{\nu}}
 =  -\lambda^{\nu}
\frac{ m_z^{\nu}}{\widetilde{\chi}_{\nu,||}}-\lambda_{e}\left(\frac{ m_z^{\nu}}{\widetilde{\chi}_{\nu,||}}
+ \frac{ m_z^{\kappa}}{\widetilde{\chi}_{\kappa,||}}\right)
\label{eq:baryaktartype2}
\end{eqnarray}
%=====================================================================================
where $m_z^{\nu}$ is the absolute value of the $z$-component of the magnetization in the sub-lattice $\nu$ and we explicitly considered that the sign of $z$-components is opposite for the sublattice $\nu$ and $\kappa$.
In the same limit,  considering $m_{\tiem{TM(RE)}}=m_{e,\tiem{TM(RE)}}+\delta m_{\tiem{TM(RE)}}$,
and following Eq. \eqref{eq:ferrilongitudinalmotionEq}
the LLB equation takes a similar form:

%========================================================================================
\begin{eqnarray}
\frac{\dot{m}_z^{\nu}}{\gamma_{\nu}}
 =  -\alpha_{\|}^{\nu}
\frac{ m_z^{\nu}}{\widetilde{\chi}_{\nu,||}}-\alpha^{\nu}_{\|}\frac{|J_{0,\nu\kappa}|}{\mu_\nu}
\left(\frac{\widetilde{\chi}_{\kappa,||}}{\widetilde{\chi}_{\nu,||}} m_z^{\nu}
-  m_z^{\kappa}\right)
\label{eq:LLBbaryaktartype2}
\end{eqnarray}
%=====================================================================================
Note that for the LLB equation the contribution of the opposite sublattice
is negative while for the Baryakhtar equation it is positive.
This has important consequences in the longitudinal inter-lattice relaxation of the sub-lattices,
changing the results of Fig. \ref{fig:phasespace}.

\begin{figure}[h!]
\centering
\includegraphics[scale=0.9,angle=90]{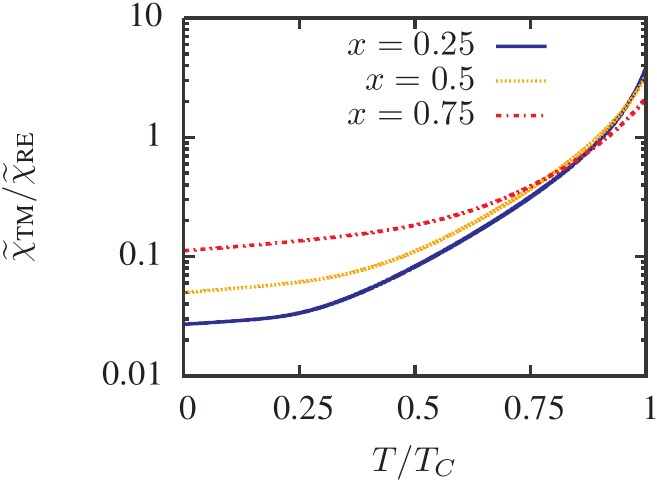}
\caption{Temperature dependence of the ratio between longitudinal susceptibilities for parameters of the GdFeCo alloy.}
\label{fig:ratiosus}
\end{figure}

In Fig. \ref{fig:ratiosus} we show the temperature dependence of the ratio of partial susceptibilities,
$\widetilde{\chi}_{\kappa,||}/\widetilde{\chi}_{\nu,||}$ appearing in Eq.\eqref{eq:LLBbaryaktartype2}.
We can see that at  temperatures not very close to $T_C$: $\widetilde{\chi}_{\tiem{TM},||}/\widetilde{\chi}_{\tiem{RE},||}\ll1$ and the contrary behavior close to $T_C$. Thus for the TM and temperatures close to $T_C$ the second term in the r.h.s. of Eq.(\ref{eq:LLBbaryaktartype2}) could be neglected and the third term with the opposite sign can compete with the first one, leading either to slowing down of the relaxation rate or even to changing its sign, as presented in Fig. \ref{fig:3cases}(a) and (b). The behavior of RE on the contrarily is dominated by this term and the sign of relaxation cannot be changed, as is seen in the same figure. Obviously this behavior cannot be described by Eq.(\ref{eq:baryaktartype2}) where all terms have the same sign. In order to have the opposite relaxation sign, one has to assume for this equation a priori that the signs of the $z$-components of magnetization in both sub-lattices are the same, i.e. to start with the ferromagnetic-like state without specifying its origin.

%====================== F_{\nu} function============
%=========================================================================================

Finally, we would like to note that because we have treated the
spin-spin interaction in MFA we have lost correlation contribution.
Consequently, both LLB and Baryakhtar equations do not describe
the energy transfer  from the uniform modes  into nonlinear spin waves and vice versa.
In ferromagnets \cite{GaraninPRB2009}
this contribution is usually two or three
orders of magnitude smaller than the contribution to relaxation through the coupling to the bath.
At this stage we do not know how large  this contribution can be in ferrimagnets.
In Ref. \cite{GaraninPRB2009} the contribution of nonlinear spin waves was artificially incremented by using a random anisotropy  to
cause non-coliniarities. In principle, in ferrimagnets  one can see a
small amount of RE as precursor of non-coliniarities,  with the strength of the order of interlattice
exchange parameter $J_{\tiem{TM-RE}}$.
For completeness, a microscopic treatment of the spin wave contribution would be desirable,
we let this task for the future.

\section{Conclusions}

We have derived the Landau-Lifshitz-Bloch equation for a two-sublattice system such as a
GdFeCo ferrimagnet for which an ultrafast switching has been reported \cite{StanciuPRL2007,RaduNATURE2011}.
Although in our derivation we refer to a TM-RE alloy,  it is equally valid for a two-component ferromagnet, as well as for an
 antiferromagnet. The generalization to more components is straightforward.
 The new equation constitutes an important step forward in classical description of the dynamics of ferrimagnets which is traditionally based
on two-coupled macroscopic LLG equations. For example, the  FMR and exchange modes have recently attracted  attention due to possibility
to optically excite them \cite{StanciuPRB2006,Mekonnen}. Their temperature dependence can be now
correctly understood in terms of our  approach \cite{Schlickeiser}.
Furthermore, recent ultrafast dynamics experiments using
XMCD showed different sublattice dynamics on ultrafast timescale in a two-sublattice magnets such as GdFeCo \cite{RaduNATURE2011}
or FeNi \cite{Raduunpub},  which can be modeled using this new approach.
Finally, this equation can serve in the future as a basis for multiscale modeling in two-component systems at
high temperatures and/or ultrafast timescales,
the same way as the LLB equation for ferromagnets \cite{KazantsevaPRB2008}.
This also opens a possibility for micromagnetic modeling of  ultrafast dynamics in large structures,
such as sub-micron and micron-size ferrimagnetic dots, whose dimensions do not allow modeling by atomistic approach.
Similarly, it will be useful for static micromagnetic modeling at high temperatures, such as thermally-driven domain wall motion in nanostructures.

The LLB equation correctly shows the possibility to reverse the sign of relaxation at high temperatures
 and, therefore, is consistent with the existence of a recently
reported ferromagnetic state in a ferrimagnet \cite{RaduNATURE2011}. The validity of the approach has been checked against full-scale atomistic simulations presented in Fig. \ref{fig:3cases}.
 However, unlike the equation,
derived by Baraykhtar and used recently to describe the GdFeCo switching \cite{MentinkPRL2012},
it is not based on the separation of timescales and on the Onsager principle.
Instead, both the  coupling to the external bath and the exchange interaction form part of the same longitudinal and transverse relaxation terms.
We show important differences in the resulting form of the equation.

Unfortunately, at the present time the compact derivation was possible only under some assumptions.
 The employed conditions certainly allow to describe
the normal modes such as ferromagnetic resonance and antiferromagnetic exchange precessional modes in ferrimagnets \cite{Schlickeiser}.
The same way the approximation is sufficient to describe the switching of ferrimagnet if
it occurs through a linear reversal path \cite{Vahaplar,MentinkPRL2012} or if sublattices
non-collinearities are not too large. Weather the applied approximation  completely describes
the situation of the ultrafast reversal  is an open question which we will investigate in the future.
For modeling, the initial paramagnetic equation (\ref{eq:LLBequationFerrimagnet}) with the MFA field \eqref{eq:HmfaRE}
and \eqref{eq:HmfaTM} can always be used, providing the check for the approximation.
Finally, up to now we were not able to derive a compact expression for the equation above
$T_C                                                                                                                                                                                                                                                                                                                                               $ which is also a necessary step for the full modeling of the ultrafast switching.

\section{Acknowledgement}

We gratefully acknowledge funding by the Spanish
Ministry of Science and Innovation under the grant
FIS2010-20979-C02-02.

\appendix
\section{}

In this appendix we present detailed derivation of Eq. \eqref{eq:NewM04-1}. We start from Eq.(\ref{eq:m0xi0}):
%=====================================================
\begin{equation}
\mathbf{m}_{0,\nu}=B(\xi_{0,\nu})\mathbf{\hat{u}}_{\nu},
\ \ \boldsymbol{\xi}_{0,\nu}\equiv\beta\mu_{\nu}\left\langle \mathbf{H}_{\nu}^{\text{MFA}}\right\rangle^{\textrm{\tiem{conf}}},
\label{eq:A0}
\end{equation}
%======================================================
where $\mathbf{\hat{u}}_{\nu}=\boldsymbol{\xi}_{0,\nu}/\xi_{0,\nu}$ and $\left\langle \mathbf{H}_{\nu}^{\text{MFA}}
\right\rangle^{\textrm{\tiem{conf}}}=\mathbf{H}_{\textrm{EX},\nu}^{\Vert}+\mathbf{H}''_{\textrm{eff},\nu}$. Here $\mathbf{H}''_{\textrm{eff}} $ contains the anisotropy, applied and the perpendicular component of the exchange field (see section III.A). In the case of a strong homogeneous exchange
field $\left|\mathbf{H}_{\textrm{EX},\nu}^{\Vert}\right|\gg\left|\mathbf{H}''_{\textrm{eff},\nu}\right|$ the MFA field
can be expanded up to first order in $H'_{\textrm{eff},\nu}$ as
%========================================================================================
\begin{eqnarray}
\left|\left\langle \mathbf{H}_{\nu}^{\text{MFA}}\right\rangle ^{\mathrm{conf}}\right|
 & \simeq & H_{\textrm{EX},\nu}^{\Vert}
+\frac{\mathbf{H}_{\textrm{EX},\nu}^{\Vert}\cdot
\mathbf{H}''_{\textrm{eff},\nu}}{\left(H_{\textrm{EX},\nu}^{\Vert}\right)}
\label{eq:A1}
\end{eqnarray}
%========================================================================================
 Therefore, $\xi_{0,\nu}=\beta\mu_{\nu}\left|\left\langle \mathbf{H}_{\nu}^{\text{MFA}}\right\rangle ^{\mathrm{conf}}\right|$ can be written
as $\xi_{0,\nu}=\xi_{\textrm{EX},\nu}+\delta\xi_{\nu}$ with $\xi_{\textrm{EX},\nu}\gg\delta\xi_{\nu}$, where we identify
$\xi_{\textrm{EX},\nu}=\beta\mu_{\nu}H_{\textrm{EX},\nu}^{\Vert}$ and
$\delta\xi_{\nu}=\beta\mu_{\nu}\left(\mathbf{H}_{\textrm{EX},\nu}^{\Vert}\cdot\mathbf{H}''_{\textrm{eff},\nu}\right)/H_{\textrm{EX},\nu}^{\Vert}$.
Expanding the Langevin function around $\xi_{\textrm{EX},\nu}$ we get
%========================================================================================
\begin{equation}
B\left(\xi_{0,\nu}\right) \simeq B_{\nu} + B'_{\nu}\delta\xi_{\nu}
\label{eq:A2}
\end{equation}
%========================================================================================
and
%========================================================================================
\begin{eqnarray}
\mathbf{\hat{u}}_{\nu}
 & \simeq & \frac{\mathbf{H}_{\textrm{EX},\nu}^{\Vert}+\mathbf{H}''_{\textrm{eff},\nu}}{H_{\textrm{EX},\nu}^{\Vert}}
\left(1-\frac{\mathbf{H}_{\textrm{EX},\nu}^{\Vert}\cdot\mathbf{H}''_{\textrm{eff},\nu}}{\left(H_{\textrm{EX},\nu}^{\Vert}\right)^{2}}\right),
\label{eq:A3}
\end{eqnarray}
%========================================================================================
where $B_{\nu}=B\left(\xi_{\textrm{EX},\nu}\right)$ and $B'_{\nu}=B'\left(\xi_{\textrm{EX},\nu}\right)$.
Substituting Eq. \eqref{eq:A2} and Eq. \eqref{eq:A3} in Eq. \eqref{eq:A0} and neglecting the terms quadratic
in ${H}''_{\textrm{eff},\nu}/|H_{\textrm{EX},\nu}^{\Vert}|$ we get
%========================================================================================
\begin{eqnarray}
\mathbf{m}_{0,\nu}
 & \simeq & B_{\nu}
\left[\frac{\mathbf{H}_{\textrm{EX},\nu}^{\Vert}+\mathbf{H}''_{\textrm{eff},\nu}}{H_{\textrm{EX},\nu}^{\Vert}}-
\frac{\left(\mathbf{H}_{\textrm{EX},\nu}^{\Vert}\cdot\mathbf{H}''_{\textrm{eff},\nu}\right)\mathbf{H}_{\textrm{EX},\nu}^{\Vert}}
{\left(H_{\textrm{EX},\nu}^{\Vert}\right)^{3}}\right]\nonumber\\
 & + & B'_{\nu}\beta\mu_{\nu}\frac{\left(\mathbf{H}_{\textrm{EX},\nu}^{\Vert}\cdot\mathbf{H}''_{\textrm{eff},\nu}\right)
\mathbf{H}_{\textrm{EX},\nu}^{\Vert}}{\left(H_{\textrm{EX},\nu}^{\Vert}\right)^{2}}.
\label{m0nu}
\end{eqnarray}
%========================================================================================
Using   the vector calculus identity
$\left(\mathbf{a}\times\mathbf{b}\right)\times\mathbf{c}=\mathbf{b}\left(\mathbf{a}\cdot\mathbf{c}\right)-
\mathbf{a}\left(\mathbf{b}\cdot\mathbf{c}\right)$  Eq.(\ref{m0nu}) can be written as
%========================================================================================
\begin{eqnarray}
\mathbf{m}_{0,\nu}
 & \simeq & B_{\nu}\frac{\mathbf{H}_{\textrm{EX},\nu}^{\Vert}}{H_{\textrm{EX},\nu}^{\Vert}}+
B'_{\nu}\beta\mu_{\nu}\frac{\left(\mathbf{H}_{\textrm{EX},\nu}^{\Vert}\cdot\mathbf{H}''_{\textrm{eff},\nu}\right)\mathbf{H}_{\textrm{EX},\nu}^{\Vert}}
{\left(H_{\textrm{EX},\nu}^{\Vert}\right)^{2}}\nonumber\\
 & - & \frac{B_{\nu}}{H_{\textrm{EX},\nu}^{\Vert}}\frac{[\left[\mathbf{H}''_{\textrm{eff},\nu}\times\mathbf{H}_{\textrm{EX},\nu}^{\Vert}\right]
\times\mathbf{H}_{\textrm{EX},\nu}^{\Vert}]]}{\left(H_{\textrm{EX},\nu}^{\Vert}\right)^{2}}.
\label{eq:A6}
\end{eqnarray}
%========================================================================================
Finally, we use $\mathbf{H}_{\textrm{EX},\nu}^{\Vert}=\left(\widetilde{J}_{0,\nu}/\mu_{\nu}\right)\mathbf{m}_{\nu}$
[see Eq. \eqref{eq:Heff}] in Eq. \eqref{eq:A6} and obtain Eq. \eqref{eq:NewM04-1}
%========================================================================================
\begin{eqnarray*}
\mathbf{m}_{0,\nu}
 & \simeq & \frac{B_{\nu}}{m_{\nu}}\mathbf{m}_{\nu}+B'_{\nu}\beta\mu_{\nu}\frac{\left(\mathbf{m}_{\nu}\cdot\mathbf{H}''_{\textrm{eff},\nu}\right)
\mathbf{m}_{\nu}}{m_{\nu}^{2}}\\
 & - &\frac{B_{\nu}\mu_{\nu}}{m_{\nu}\widetilde{J}_{0,\nu}}\frac{\left[\left[\mathbf{H}''_{\textrm{eff},\nu}\times\mathbf{m}_{\nu}\right]
\times\mathbf{m}_{\nu}\right]}{m_{\nu}^{2}}.
\end{eqnarray*}
%========================================================================================

\end{document}